\documentclass{ws-rv961x669}
\usepackage[square]{ws-rv-van}
\usepackage{ws-rv-thm}     
\usepackage{subfigure}     
\usepackage{bbm}
\makeindex

\begin{document}

\setcounter{chapter}{-1}

\chapter[Extended anti-de Sitter Hypergravity  in $2+1$ Dimensions]{Extended anti-de Sitter Hypergravity  in $2+1$ Dimensions and Hypersymmetry Bounds}

\author[M. Henneaux, A P\'erez, D. Tempo and R. Troncoso]{Marc Henneaux$^{a}$, Alfredo P\'erez$^{b}$, David Tempo$^{a,b}$ and Ricardo Troncoso$^{b}$}

\address{$^a$Universit\'{e} Libre de Bruxelles and International Solvay Institutes, \\
ULB Campus Plaine C.P.231, B-1050 Bruxelles, Belgium \\
$^b$Centro de Estudios Cient\'{i}ficos (CECs), Av. Arturo Prat 514, Valdivia,
Chile}

\begin{abstract}
In a recent paper (JHEP {\bf 1508} (2015) 021), we have investigated hypersymmetry bounds in the context of simple anti-de Sitter hypergravity in $2+1$ dimensions. We showed that these bounds involved non linearly the spin-$2$ and spin-$4$ charges, and were saturated by a class of extremal black holes, which are $\frac14$-hypersymmetric. We continue the analysis here by considering $(M,N)$-extended anti-de Sitter hypergravity models, based on the superalgebra $osp(M \vert 4) \oplus osp(N \vert 4)$. The asymptotic symmetry superalgebra is then the direct sum of two-copies of a $W$-superalgebra that contains $so(M)$ (or $so(N)$) Kac-Moody currents of conformal weight $1$, fermionic generators of conformal weight $5/2$ and bosonic generators of conformal weight $4$ in addition to the Virasoro generators. The nonlinear hypersymmetry bounds on the conserved charges are derived and shown to be saturated by a class of extreme hypersymmetric black holes which we explicitly construct.
\end{abstract}


\body


\section{Introduction\label{sec:Introduction}}

Simple anti-de Sitter hypergravity in three dimensions is a consistent
higher spin field theory involving fields of spins $2$, $4$ and
$\frac{5}{2}$ that is invariant under hypersymmetry, a higher spin
fermionic symmetry with spin-$\frac{3}{2}$ parameter. In the limit
of zero cosmological constant, the spin-$4$ field decouples and the
theory of the remaining fields reduces to the hypergravity model of
\cite{Aragone:1983sz} that has been recently reformulated as a Chern-Simons
theory in \cite{Fuentealba:2015jma}. The theory has no local degrees
of freedom, but possesses the rich asymptotics of higher spin gauge
fields in $2+1$ dimensions described by $W$-(super)algebras \cite{Henneaux-HS,Theisen-HS,Henneaux:2012ny},
in this case $W_{\left(2,\frac{5}{2},4\right)}$ \cite{Henneaux:2015ywa}.

The asymptotic symmetry algebra has interesting consequences since
it implies ``hypersymmetry bounds'', in much the same way as supersymmetry
implies supersymmetry bounds. In \cite{Henneaux:2015ywa} we explicitly
derived the hypersymmetry bounds for simple AdS hypergravity and analyzed
them for different types of solutions, in particular, for black holes.
We showed that the hypersymmetric black holes saturate the bounds
and are extremal, in the sense that they lie on the border of the
region within which a sensible thermodynamics (real entropy) can be
defined.

The purpose of this work is to extend the analysis to hypergravity
models with more hypersymmetries (``extended hypergravity''). This
is done along the following lines. First, in the next section (section \ref{sec:Extended}), we describe the extended
hypergravity models: we discuss the underlying superalgebras and write down the action. Then, in Section \ref{sec:Asymptotic}, we study the asymptotics using well-established methods
and show that the asymptotic superalgebra is an extension of the algebra
of \cite{Henneaux:2015ywa} by $so(M)$ (respectively $so(N)$) Kac-Moody currents under which
the fermionic hypercharges transform in the $\underbar{{\bf {M}}}$ (respectively, in the $\underbar{{\bf {N}}}$).
We derive the form of the nonlinear hypersymmetry bounds in \ref{sec:HBounds}. Next,
in Section \ref{sec:Black-holes}, we construct the black hole solutions
and discuss their thermodynamics. Finally, we show that the hypersymmetric
black holes are extremal and saturate the hypersymmetry bounds (Section
\ref{sec:Killing-vector-spinor-equation}). Section \ref{sec:Conclusions}
collects our concluding remarks.

\section{Extended anti-de Sitter hypergravities in $2+1$ dimensions}

\label{sec:Extended} \setcounter{equation}{0}

$(M,N)$-extended AdS hypergravities in three-dimensional spacetimes
are hypersymmetric extensions of $sp(4)$ higher spin gravity, described
by a Chern-Simons theory with gauge algebra $sp(4)\oplus sp(4)$ where
the gravitational subalgebra $sp(2)\simeq sl(2,\mathbb{R})$ is principally
embedded on each side. The $sp(4)$ higher spin gravity contains the
graviton and a spin-$4$ field. Here, the word ``spin'' refers to
the conformal weight of the corresponding asymptotic generators in
the conformal algebra at infinity, see below.

One may ``hypersymmetrize'' the $sp(4)$ higher spin gravity and
construct $(M,N)$-extended AdS hypergravities, with $M$ hypersymmetries
in one chiral sector and $N$ hypersymmetries in the other, by embedding
$sp(4)\oplus sp(4)$ in an appropriate superalgebra. We consider hypersymmetric
extensions such that the resulting superalgebra has the following
properties: 
\begin{itemize}
\item The bosonic subalgebra is the direct sum $sp(4)\oplus{\mathcal{G}}$
where ${\mathcal{G}}$ is a compact algebra. 
\item The fermionic generators transform in the $\underbar{{\bf 4}}$ of
$sp(4)$ (and in some  representation of ${\mathcal{G}}$ on which we do not impose any a priori requirement). 
\end{itemize}
The first condition implies that the extra bosonic fields in the theory,
coming in addition to the graviton and its spin-$4$ companion, have
all lower spin $1$. These extra fields are just the gauge fields
associated with the internal $R$-symmetry described by ${\mathcal{G}}$.
The second condition guarantees that the fermionic fields have all
spin $\frac{5}{2}$ (``hypergravitini'').

The algebra $sp(4)$ is the algebra underlying anti-de Sitter hypergravity
in three dimensions, but it is also the anti-de Sitter algebra in
$4$ dimensions. As such, its graded extensions have been systematically
studied in the early days of supergravity. It turns out that there
is only one class of graded extensions, given by $osp(M\vert4)$ \cite{Kac,Nahm:1977tg}.
Thus, while there are seven distinct types of extended supergravity
models in three dimensions \cite{Kac,Nahm:1977tg,Gunaydin:1986fe,AchucarroTownsend1,AchucarroTownsend2},
there is only one type of extended hypergravity models. The underlying
superalgebras are $osp(M\vert4)\oplus osp(N\vert4)$.

The $(M,N)=(1,1)$ case of \cite{Henneaux:2015ywa} is described by
the superalgebra $osp(1\vert4)\oplus osp(1\vert4)$. It contains,
in addition to the graviton and its spin-$4$ companion, a spin $\frac{5}{2}$ field on each chiral side, called the ``hypergravitino".  In the extended case, there are more ``hypergravitini" and these transform in the   $\underbar{{\bf {M}}}$ (respectively, in the $\underbar{{\bf {N}}}$) of $so(M)$ (respectively, $so(N)$).  There are also extra gauge fields transforming in the adjoint of $so(M)$ (respectively, $so(N)$).

For definiteness, we shall focus from now on the chiral sector with
superalgebra $osp(M\vert4)$. Similar considerations apply to the
other sector. The (anti)commutation relations of $osp(M\vert4)$ are
explicitly: 
\begin{align}
\left[L_{i},L_{j}\right] & =\left(i-j\right)L_{i+j}\thinspace,\nonumber \\
\left[L_{i},U_{m}\right] & =\left(3i-m\right)U_{i+m}\thinspace,\nonumber \\
\left[L_{i},T^{IJ}\right] & =0\thinspace,\nonumber \\
\left[L_{i},\mathcal{S}_{p}^{I}\right] & =\left(\frac{3}{2}i-p\right)\mathcal{S}_{i+p}^{I}\thinspace,\nonumber \\
\left[U_{m},U_{n}\right] & = \nonumber \\
& \hspace{-1.2cm} \frac{1}{12}\left(m-n\right)\left(\left(m^{2}+n^{2}-4\right)\left(m^{2}+n^{2}-\frac{2}{3}mn-9\right)-\frac{2}{3}\left(mn-6\right)mn\right)L_{m+n}\nonumber \\
 & +\frac{1}{6}\left(m-n\right)\left(m^{2}-mn+n^{2}-7\right)U_{m+n}\thinspace,\label{eq:WedgeAlgebra}\\
\left[T^{IJ},T^{KL}\right] & =\delta^{IK}T^{JL}-\delta^{IL}T^{JK}-\delta^{JK}T^{IL}+\delta^{JL}T^{IK}\thinspace,\nonumber \\
\left[U_{m},T^{IJ}\right] & =0\thinspace,\nonumber \\
\left[U_{m},\mathcal{S}_{p}^{I}\right] & =\frac{1}{24}\left(2m^{3}-8m^{2}p+20mp^{2}+82p-23m-40p^{3}\right)\mathcal{S}_{i+p}^{I}\thinspace,\nonumber \\
\left[T^{IJ},\mathcal{S}_{p}^{K}\right] & =\delta^{IK}\mathcal{S}_{p}^{J}-\delta^{JK}\mathcal{S}_{p}^{I}\thinspace,\nonumber \\
\left\{ \mathcal{S}_{p}^{I},\mathcal{S}_{q}^{J}\right\}  & =\delta^{IJ}\left(U_{p+q}+\frac{1}{12}\left(6p^{2}-8pq+6q^{2}-9\right)L_{p+q}\right) \nonumber \\
&-\frac{5}{12}(p-q)(2p^{2}+2q^{2}-5)T^{IJ}\thinspace.\nonumber 
\end{align}
Here $L_{i}$, with $i=0,\pm1$, stand for the generators that
span the gravitational $sl\left(2,\mathbb{R}\right)$ subalgebra, while $T^{IJ}=-T^{JI}$,
with $I,J=1,\cdots,M$, are the spin-0 $so(M)$ generators, which will yield spin-1 fields in the Chern-Simons theory. The $U_{m}$
and ${\cal S}_{p}^{I}$, with $m=0,\pm1,\pm2,\pm3$ and $p=\pm\frac{1}{2},\pm\frac{3}{2}$,
will yield the spin-4 and spin-$\frac{5}{2}$ fields,
respectively.

The dynamics of $(M,N)$-extended hypergravity follows from the difference
of two Chern-Simons actions, $I=I_{CS}\left[A^{+}\right]-I_{CS}\left[A^{-}\right]$,
with 
\begin{equation}
I_{CS}\left[A\right]=\frac{k_{4}}{4\pi}\int str\left[AdA+\frac{2}{3}A^{3}\right]\thinspace,\label{eq:ICS}
\end{equation}
where the level, $k_{4}=k/10$, is expressed in terms of the Newton
constant and the AdS radius according to $k=\ell/4G$. In eq. (\ref{eq:ICS})
$str\left[\cdots\right]$ stands for the supertrace of the fundamental
($(4+M)\times(4+M)$ or $(4+N)\times(4+N)$) matrix representation
of $osp(M\vert4)$ and the gauge fields $A^{\pm}$ correspond to the
two independent copies $osp(M\vert4)$ and $osp(N\vert4)$. A convenient
matrix representation of the generators $T^{IJ}$ is such that the
lower diagonal block is given by 
\[
\left(T^{IJ}\right)_{\;\;L}^{K}=-2\delta^{K[I}\delta_{L}^{J]}\;,
\]
and hence 
\[
str\left(T^{IJ}T^{KL}\right)=4\delta^{K[I}\delta^{J]L}\;.
\]

\section{Asymptotic structure of extended hypergravities\label{sec:Asymptotic} }

\subsection{Boundary conditions}

In order to discuss the boundary conditions, we perform -- as it has now become standard --  the gauge transformation of \cite{CHvD} that eliminates asymptotically
the radial dependence of the connections, so that $A^{\pm}=g_{\pm}^{-1}a^{\pm}g_{\pm}+g_{\pm}^{-1}dg_{\pm}$,
with 
\begin{equation}
a^{\pm}=a_{\varphi}^{\pm}\left(t,\varphi\right)d\varphi+a_{t}^{\pm}\left(t,\varphi\right)dt\thinspace
\end{equation}
(to leading order). Then, following the lines of \cite{Henneaux-HS,Theisen-HS,HASugra,Henneaux:2012ny},
we impose that at any fixed time slice $t=t_{0}$, the deviations
with respect to the reference background go asymptotically along the
lowest (highest) $sl(2)$-weight vectors for each $sl(2)$-representation occurring in the theory, i.e., 
\begin{align}
a_{\varphi}^{\pm} & =L_{\pm1}-\frac{2\pi}{k}\tilde{\mathcal{L}}^{\pm}\left(\varphi\right)L_{\mp1}+\frac{\pi}{5k}\mathcal{U}^{\pm}\left(\varphi\right)U_{\mp3}-\frac{2\pi}{k}\psi_{I}^{\pm}\left(\varphi\right)\mathcal{S}_{\mp\frac{3}{2}}^{I}-\frac{5\pi}{k}\mathcal{J}_{IJ}^{\pm}(\varphi)T^{IJ}\thinspace.\label{eq:atheta}
\end{align}
All components $\mathcal{J}_{IJ}^{\pm}$ along the internal symmetry generators $T^{IJ}$, which are $sl(2)$-scalars, are allowed.  In (\ref{eq:atheta}), $\tilde{\mathcal{L}}^{\pm}\left(\varphi\right)$ is defined
in terms of what will become the Virasoro generators $\mathcal{L}^{\pm}$
through 
\begin{equation}
\tilde{\mathcal{L}}^{\pm}=\mathcal{L}^{\pm}-\frac{5\pi}{2k}{\cal J}_{IJ}^{\pm}{\cal J}^{\pm IJ}\;.
\end{equation}
The two expressions differ by the familiar Sugawara term quadratic
in the currents.

\subsection{Asymptotic symmetries}

\label{Subsec:Asy}

Exactly as in \cite{Henneaux-HS,Theisen-HS,HASugra,Henneaux:2012ny},
one then finds that the fall-off conditions (\ref{eq:atheta}) are maintained under a
restricted set of gauge transformations, $\delta a^{\pm}=d\Omega^{\pm}+\left[a^{\pm},\Omega^{\pm}\right]$,
where, on each slice, the Lie-algebra-valued parameters 
\begin{equation}
\Omega^{\pm}=\Omega^{\pm}\left[\epsilon_{\pm},\chi_{\pm},\zeta_{IJ\pm},\vartheta_{\pm}^{I}\right]\thinspace,\label{eq:Omegamn}
\end{equation}
depend on ($2+\frac{M^{\pm}(M^{\pm}-1)}{2}$) bosonic and $M^{\pm}$
fermionic functions of $\varphi$, given by $\epsilon_{\pm},\chi_{\pm},\zeta_{IJ\pm}$,
and $\vartheta_{\pm}$, respectively. Here, we have set $M^{+}=M$
and $M^{-}=N$. They take the form 
\begin{eqnarray}
\Omega^{\pm}\left[\epsilon_{\pm},\chi_{\pm},\zeta_{IJ\pm},\vartheta_{\pm}^{I}\right] & = & \epsilon_{\pm}\left(\varphi\right)L_{\pm1}-\chi_{\pm}\left(\varphi\right)U_{\pm3}\mp\vartheta_{\pm\left[I\right]}\left(\varphi\right)\mathcal{S}_{\pm\frac{3}{2}}^{\left[I\right]}\nonumber \\
 &  & \hspace{-1.5cm}+\left(\zeta_{IJ\pm}\left(\varphi\right)-\frac{5\pi}{k}\epsilon_{\pm}\left(\varphi\right){\cal J}_{IJ}^{\pm}\right)T^{IJ}+\eta^{\pm}\left[\epsilon_{\pm},\chi_{\pm},\nu_{IJ\pm},\vartheta_{\pm}^{I}\right]\thinspace,\label{eq:Omegamp}
\end{eqnarray}
where the $\eta^{\pm}$'s, and the precise way in which the fields ${\cal L}^{\pm}$,
${\cal U}^{\pm}$, $\mathcal{J}_{IJ}^{\pm}$, $\psi^{\pm I}$ transform,
are explicitly given in Appendix \textbf{\ref{Appendix:Asy}}.  These expressions involve the fields  ${\cal L}^{\pm}$,
${\cal U}^{\pm}$, $\mathcal{J}_{IJ}^{\pm}$, $\psi^{\pm I}$ and the independent gauge parameters $\epsilon_{\pm},\chi_{\pm},\zeta_{IJ\pm}$,
and $\vartheta_{\pm}$, as well as their derivatives with respect to $\varphi$.

The boundary conditions (\ref{eq:atheta}) define phase space at a given instant of time. Phase space histories fulfill (\ref{eq:atheta}) at all times, i.e., take the form (\ref{eq:atheta}) with the functions $\tilde{\mathcal{L}}^\pm$, $\mathcal{U}^\pm$, $\psi_I^\pm$ and $\mathcal{J}_{IJ}^\pm$ now depending also on $t$.  These boundary conditions are of course preserved by gauge transformations of the form (\ref{eq:Omegamp}) with
parameters $\epsilon_{\pm}\left(t,\varphi\right)$, $\chi_{\pm}\left(t,\varphi\right)$, $\vartheta_{\pm\left[I\right]}\left(t,\varphi\right)$, $\zeta_{IJ\pm}\left(t,\varphi\right)$  that are time-dependent too.  In particular, the motion in time is a gauge transformation with gauge parameter $a_t^\pm$.  This implies that
 the asymptotic
behaviour of $a_{t}^{\pm}$ has to be given by \cite{HPTT,Bunster:2014mua}
\begin{equation}
a_{t}^{\pm}=\pm\Omega^{\pm}\left[\xi_{\pm},\mu_{\pm},\nu_{IJ\pm},\varrho_{\pm}^{I}\right]\;,\label{eq:at}
\end{equation}
where $\Omega^{\pm}$ is defined through (\ref{eq:Omegamp}), and
$\xi_{\pm}$, $\mu_{\pm}$, $\nu_{IJ\pm}$, $\varrho_{\pm}^{I}$ can be identified with
the \textquotedblleft chemical potentials\textquotedblright   $\,$ when one goes to the thermodynamical formulation.  Once the temporal components of the vector potential have been chosen, the  parameters $\epsilon_{\pm}$,
$\chi_{\pm}$, $\zeta_{IJ\pm}$, $\vartheta_{\pm}^{I}$ of the residual gauge transformations must fulfill certain
differential equations of first order in time expressing that the $a_t^\pm$'s are left invariant by the transformations, which may be regarded
as \textquotedblleft deformed chirality conditions\textquotedblright .

\subsection{Generators of asymptotic symmetries}

Following the canonical approach \cite{Regge:1974zd}, one finds that
the generators of the asymptotic symmetries are 
\begin{equation}
{\cal Q}^{\pm}\left[\epsilon_{\pm},\chi_{\pm},\vartheta_{I\pm}\right]=-\int d\varphi\left(\epsilon_{\pm}{\cal L}^{\pm}+\chi_{\pm}{\cal U}^{\pm}-\zeta_{\pm}^{IJ}{\cal J}_{IJ}^{\pm}-i\vartheta_{\pm}^{\;I}\psi_{I}^{\pm}\right)\thinspace.
\end{equation}
(modulo bulk terms proportional to the constraints that we will not
write explicitly and that can be taken strongly equal to zero if one
uses the Dirac bracket - which coincides with the Poisson bracket
for gauge invariant functions).

Since the Poisson brackets fulfill $\left[{\cal Q}\left[\eta_{1}\right],{\cal Q}\left[\eta_{2}\right]\right]_{PB}=-\delta_{\eta_{1}}{\cal Q}\left[\eta_{2}\right]$,
the algebra of the canonical generators can be easily found from the
transformation law of the fields, and it is explicitly written down in
Appendix \textbf{\ref{Appendix:PB}}.

Expanding in Fourier modes, $X=\frac{1}{2\pi}\sum_{m}X_{m}e^{im\varphi}$,
the asymptotic symmetry algebra reads
\begin{align*}
i\left[L_{m},L_{n}\right]_{PB} & =\left(m-n\right)L_{m+n}+\frac{k}{2}m^{3}\delta_{m+n}^{0}\;,\\
i\left[L_{m},U_{n}\right]_{PB} & =\left(3m-n\right)U_{m+n}\;,\\
i\left[L_{m},J_{n}^{IJ}\right]_{PB} & =-nJ_{m+n}^{IJ}\;,\\
i\left[L_{m},\psi_{n}^{I}\right]_{PB} & =\left(\frac{3}{2}m-n\right)\psi_{m+n}^{I}\;,
\end{align*}
\begin{align*}
i\left[U_{m},J_{n}^{IJ}\right]_{PB} & =0\;,\\
i\left[J_{m}^{IJ},\left(J_{KL}\right)_{n}\right]_{PB} & =-4i\delta_{\;\left[K\right.}^{\left[I\right.}\left({\cal J}_{\;\;\left.L\right]}^{\left.J\right]}\right)_{m+n}-\frac{2k}{5}n\delta_{\;\;K}^{\left[I\right.}\delta_{\;\;L}^{\left.J\right]}\delta_{m+n}^{0}\;,\\
\left[J_{m}^{IJ},\psi_{n}^{K}\right]_{PB} & =2\delta^{K\left[I\right.}\psi_{m+n}^{\left.J\right]}\;,
\end{align*}
\begin{align}
i\left[U_{m},U_{n}\right]_{PB} & =\frac{1}{2^{2}3^{2}}\left(m-n\right)\left(3m^{4}-2m^{3}n+4m^{2}n^{2}-2mn^{3}+3n^{4}\right)\left(L_{m+n}-\frac{5\pi}{2k}\Lambda_{m+n}^{\left(2\right)}\right)\nonumber \\
 & +\frac{1}{6}\left(m-n\right)\left(m^{2}-mn+n^{2}\right){\cal U}_{m+n}-\frac{2^{3}3\pi}{k}\left(m-n\right)\Lambda_{m+n}^{\left(6\right)}\nonumber \\
 & -\frac{7^{2}\pi}{3^{2}k}(m-n)\left(m^{2}+4mn+n^{2}\right)\tilde{\Lambda}_{m+n}^{\left(4\right)}+\frac{k}{2^{3}3^{2}}m^{7}\delta_{m+n}^{0}\;,\nonumber \\
i\left[U_{m},\psi_{n}^{I}\right]_{PB} & =\frac{1}{2^{2}3}\left(m^{3}-4m^{2}n+10mn^{2}-20n^{3}\right)\psi_{m+n}^{I}-\frac{23\pi}{3k}i\Lambda_{m+n}^{\left(11/2\right)I}\nonumber \\
 & +\frac{70\pi}{k}im^{2}\Lambda_{m+n}^{\left(7/2\right)I}+\frac{\pi}{3k}\left(23m-82n\right)\Lambda_{m+n}^{\left(9/2\right)I}\;,\nonumber \\
i\left[\psi_{m}^{I},\psi_{n}^{J}\right]_{PB} & =\delta^{IJ}\left[U_{m+n}+\frac{1}{2}\left(m^{2}-\frac{4}{3}mn+n^{2}\right)\left(L_{m+n}-\frac{5\pi}{2k}\Lambda_{m+n}^{\left(2\right)}\right)+\frac{k}{6}m^{4}\delta_{m+n}^{0}\right]\nonumber \\
 & -\frac{5}{6}i\left(m-n\right)\left(m^{2}+n^{2}\right)J_{m+n}^{IJ}\nonumber \\
 & -\frac{25\pi}{k}\left(m^{2}+n^{2}\right)\Lambda_{m+n}^{\left(2\right)IJ}+\frac{50\pi}{3k}i\left(m\Lambda_{m+n}^{\left(3\right)JI}+n\Lambda_{m+n}^{\left(3\right)IJ}\right)-\frac{3\pi}{k}\Lambda_{m+n}^{\left(4\right)\left(IJ\right)}\;,\label{eq:PBFermionic}
\end{align}
where $\Lambda_{m}^{\left(l\right)}$ stand for the Fourier modes
of the corresponding nonlinear terms (see Appendix \textbf{\ref{Appendix:PB}}).
The central charge in the Virasoro subalgebra is the same as that
for pure gravity \cite{Brown:1986nw}.

The modes fulfill the following reality conditions: $\left(L_{m}\right)^{*}=L_{-m}$,
$\left(U_{m}\right)^{*}=U_{-m}$, $\left(\psi_{m}^{I}\right)^{*}=\psi_{-m}^{I}$,
$\left(\mathcal{J}_{m}^{IJ}\right)^{*}=\mathcal{J}_{-m}^{IJ}$ so
that the functions ${\cal L}^{\pm}$, ${\cal U}^{\pm}$, $\psi^{\pm}$
and $\mathcal{J}^{IJ}$ are real.

The asymptotic symmetry algebra given above is the classical limit
of an extension of the superalgebra $W_{\left(2,\frac{5}{2},4\right)}$  of \cite{FigueroaO'Farrill:1991pb,Bellucci:1994xa}. This extension involves $M$ spin-$\frac{5}{2}$ fermionic generators $\psi_{m}^{I}$
and $\frac{M(M-1)}{2}$ Kac-Moody currents $\mathcal{J}_{n}^{IJ}$. The spin-$\frac{5}{2}$
fermionic generators $\psi_{m}^{I}$ are the ``hypersymmetry'' generators,
and transform in the $\underbar{{\bf M}}$ of $so(M)$.  The algebra $W_{\left(2,\frac{5}{2},4\right)}$ corresponds to $M=1$.

\section{Hypersymmetry bounds from the asymptotic symmetry algebra}
\setcounter{equation}{0}
\label{sec:HBounds}

\subsection{Boundary conditions and spectral flow}

We focus for definiteness on the $+$ copy and drop the subscript
``$+$''. Similar considerations apply to the $-$ sector.

The fermions are subject to boundary conditions of the form 
\begin{equation}
\psi_{I}(\varphi+2\pi)=R_{IJ}\psi_{J}(\varphi)
\end{equation}
where the matrix $R=(R_{IJ})$ is an element of $O(M)$, which we
can take to be either the identity, or a fixed element of $O(M)$
with determinant $-1$ discussed below. Different boundary
conditions are related to these ones by spectral flow \cite{Schwimmer:1986mf}
(see also \cite{HASugra} for a discussion in the similar AdS$_{3}$
extended supergravity context).

When $M$ is odd, one may assume $R=\mathbbm{1}$ (periodic boundary
conditions) or $R=-\mathbbm{1}$ (antiperiodic boundary conditions).
In both cases, the affine generators ${\mathcal{J}}_{IJ}$ are periodic
and the corresponding affine algebra is untwisted. When $M$ is even,
$M=2r$, one may assume $R=\mathbbm{1}$ (periodic boundary conditions)
or, if $R\not=\mathbbm{1}$, that it defines an outer automorphism
of $SO(2r)$. In that latter case, the affine generators are not periodic
and the affine algebra is twisted.

We shall restrict the analysis to periodic boundary conditions (``Ramond
case''). This is motivated by the fact that we are interested in
black holes. The situations found in $\left(1,1\right)$ hypergravity
and supergravity indicate that black hole solutions naturally admit in both cases
the periodic spin structure \cite{Henneaux:2015ywa,Coussaert:1993jp}.
Note that the antiperiodic case (``Neveu-Schwarz'' case) is automatically
included when $M$ is even since then, as mentioned above, it can
be related to the periodic case by spectral flow \cite{Schwimmer:1986mf}
($-\mathbbm{1}\in SO(2m)$).  With periodic boundary conditions, the Fourier labels $m$ in $X=\frac{1}{2\pi}\sum_{m}X_{m}e^{im\varphi}$ are integers for all fields $X$.

\subsection{Hypersymmetry bounds}

The Poisson Bracket of the fermionic generator of the asymptotic symmetry
hyperalgebra in (\ref{eq:PBFermionic}), implies interesting hypersymmetry
bounds. These were discussed in great generality in \cite{Henneaux:2015ywa}.
Here, we focus on the bounds that hold in the context of periodic
boundary conditions.

We consider bosonic configurations carrying global charges with only
zero modes, given by $L_{0}=2\pi{\cal L}$, $U_{0}=2\pi{\cal U}$
and $J_{0}^{IJ}=2\pi\mathcal{J}^{IJ}$, for each copy. Furthermore,
we assume without loss of generality that the affine Kac-Moody currents
have been brought to the Cartan subalgebra by conjugation, so that
$\mathcal{J}^{IJ}$ has only non-vanishing components $\mathcal{J}^{2i-1\,2i}$
($i=1,2,\cdots$, rank$SO(M)=\left[\frac{M}{2}\right]=r$). We set
$M=2r$ when $M$ is even, or $M=2r+1$ when $M$ is odd, and 
\begin{equation}
\mathcal{J}^{2i-1\,2i}=j_{i}\epsilon^{2i-1\,2i}
\end{equation}

The anticommutators of the hypersymmetry generators with $m=-n=p\geq0$
are then found to reduce to 
\begin{align}
\left(2\pi\right)^{-1}\left(\hat{\psi}_{p}^{I}\hat{\psi}_{-p}^{J}+\hat{\psi}_{-p}^{J}\hat{\psi}_{p}^{I}\right) & =B_{p}^{IJ}\;,
\end{align}
with 
\begin{align}
B_{p}^{IJ} & =\left({\cal U}+\frac{3\pi}{k}\tilde{{\cal L}}^{2}\right)\delta^{IJ}+\frac{500\pi^{2}}{3k^{2}}\mathcal{J}_{\;\;K}^{I}\left(\tilde{{\cal L}}\mathcal{J}^{JK}+\frac{5\pi}{k}\mathcal{J}_{MN}\mathcal{J}^{JM}\mathcal{J}^{KN}\right)\nonumber \\
 & -\frac{100i\pi}{3k}\left(\tilde{{\cal L}}\mathcal{J}^{IJ}+\frac{10\pi}{k}\mathcal{J}_{\;\;M}^{I}\mathcal{J}^{ML}\mathcal{J}_{\;\;L}^{J}\right)p+\frac{5}{3}\left(\delta^{IJ}\tilde{{\cal L}}+\frac{30\pi}{k}\mathcal{J}_{\;\;K}^{I}\mathcal{J}^{JK}\right)p^{2}\nonumber \\
 & -\frac{10}{3}i\mathcal{J}^{IJ}p^{3}+\frac{k}{12\pi}\delta^{IJ}p^{4}\;.\label{Eq:B}
\end{align}
Note that 
\begin{equation}
\left(B_{p}^{IJ}\right)^{\dagger}=B_{p}^{JI}\;,
\end{equation}
and as one sees explicitly from (\ref{Eq:B}), $B_{p}^{IJ}$ has both
a real symmetric part and a pure imaginary, antisymmetric part.

Now, the hermitian operator $\hat{\psi}_{p}^{I}\hat{\psi}_{-p}^{I}+\hat{\psi}_{-p}^{I}\hat{\psi}_{p}^{I}$
is positive definite for each $I$ and $p$. This implies, in the
classical limit, that the global charges fulfill the bounds 
\begin{equation}
B_{p}^{II}\geq0\;,
\end{equation}
(no summation over $I$). The bound $B_{0}^{II}\geq0$ for $p=0$
reads 
\begin{equation}
B_{0}^{II}\equiv\left({\cal U}+\frac{3\pi}{k}\tilde{{\cal L}}^{2}\right)+\frac{500\pi^{2}}{3k^{2}}\left(\tilde{{\cal L}}(j_{i})^{2}+\frac{5\pi}{k}(j_{i})^{4}\right)\geq0\thinspace,\label{eq:B0}
\end{equation}
with $I=1$, $2$,$\cdots$$2r$. Note that when $M$ is odd, there
is an additional bound corresponding to $I=2r+1$, 
\begin{equation}
B_{0}^{2r+1\,2r+1}\equiv\left({\cal U}+\frac{3\pi}{k}\tilde{{\cal L}}^{2}\right)\geq0\thinspace.
\end{equation}
These bounds are manifestly nonlinear.

One can express the bounds for $p>0$ in terms of the bounds for $p=0$
as 
\begin{align*}
B_{p}^{II} & =B_{0}^{II}+\frac{5}{3}\left(\tilde{{\cal L}}+\frac{30\pi}{k}(j_{i})^{2}\right)p^{2}+\frac{k}{12\pi}p^{4}\geq0\,.
\end{align*}
Now, in the black hole case, one must have $\tilde{{\cal L}}\geq0$ (see below)
and so one finds that the bounds with $p>0$ are automatic consequences
of bounds with $p=0$, which are thus the strongest.

One can derive further bounds involving the mixed terms $B_{p}^{IJ}$
with $J\not=I$. To illustrate the procedure, consider for definitess
$I=1$ and $J=2$, for which $B_{p}^{12}$ does not identically vanish.
Form the complex fields $\chi_{p}=\psi_{p}^{1}+i\psi_{p}^{2}$ and
$\omega_{p}=\psi_{p}^{1}-i\psi_{p}^{2}$. From the conditions $\chi_{p}(\chi_{p})^{\dagger}+(\chi_{p})^{\dagger}\chi_{p}\geq0$
and $\omega_{p}(\omega_{p})^{\dagger}+(\omega_{p})^{\dagger}\omega_{p}\geq0$,
one gets 
\begin{equation}
B_{p}^{11}+B_{p}^{22}\geq\pm i(B_{p}^{12}-B_{p}^{21}),
\end{equation}
i.e., given that $B_{p}^{11}=B_{p}^{22}$ and $B_{p}^{12}=-B_{p}^{21}$,
\begin{equation}
B_{p}^{11}\geq\pm iB_{p}^{12}\:. \label{eq:MixedBound}
\end{equation}
In general, this bound is independent from the previous ones, but it is not so in the black hole case.  Indeed, the condition (\ref{eq:MixedBound}) can be conveniently factorized as
\begin{equation}
\left[\left(p-\frac{10\pi}{k}j_{1}\right)^{2}+\lambda_{\left[+\right]}^{2}\right]\left[\left(p-\frac{10\pi}{k}j_{1}\right)^{2}+\lambda_{\left[-\right]}^{2}\right]\geq0\;,\label{eq:MixedB}
\end{equation}
where $\pm\lambda_{\left[+\right]}$ and $\pm\lambda_{\left[-\right]}$
correspond to the eigenvalues of the $sp\left(4\right)$ dynamical
gauge fields introduced in the next section. 
In the black hole case, these eigenvalues
are necessarily real (see below), so that for this class of solutions
the bounds in (\ref{eq:MixedB}) are clearly fulfilled.   One refers for this reason to the bounds (\ref{eq:B0}) with $p = 0$ as the ``strongest bounds" in the black hole context.

\section{Black holes\label{sec:Black-holes}}

\subsection{Black hole connection and regularity conditions}

Higher spin black holes generalizing the pure gravity black hole \cite{BTZ,BHTZ} have been investigated first in the pioneering work \cite{GK,AGKP,CM}, reviewed in \cite{Ammon:2012wc}.   A different class of black hole solutions differring in their asymptotics was subsequently derived in \cite{HPTT,Bunster:2014mua,PTTreview}.  We follow this approach as it is clearly compatible with the asymptotic $W$-symmetry algebra exhibited above.  

In the absence of a well-defined spacetime geometry, higher spin black holes are defined through the Euclidean continuation \cite{GK,AGKP}, as regular flat connections on the solid torus with well-defined thermodynamics (real entropy).  We follow this point of view but, however, as in \cite{Bunster:2014mua}, we impose the above boundary conditions on the
connection and not ones that would modify the asymptotic behaviour of $a^{pm}_\varphi$.   For the $(M,M)$-extended AdS hypergravity theory,  the Euclidean connection
that describes the black holes is a direct generalization of the simple hypergravity black hole of \cite{Henneaux:2015ywa} and can be written as 
\begin{align}
a & =\left(L_{1}-\frac{2\pi}{k}\tilde{\mathcal{L}}L_{-1}+\frac{\pi}{5k}\mathcal{U}U_{-3}-\frac{5\pi}{k}{\cal J}_{IJ}T^{IJ}\right)d\varphi-\left\{ i\xi\left(L_{1}-\frac{2\pi}{k}\tilde{\mathcal{L}}L_{-1}+\frac{\pi}{5k}\mathcal{U}U_{-3}\right.\right.\nonumber \\
 & \left.+\frac{5\pi}{k}{\cal J}_{IJ}T^{IJ}\right)-i\mu\left[U_{3}-\frac{6\pi}{k}\tilde{\mathcal{L}}U_{1}-\frac{22}{15}\frac{\pi^{2}}{k^{2}}\left(\mathcal{U}+\frac{60\pi}{11k}\tilde{\mathcal{L}}^{2}\right)\tilde{\mathcal{L}}U_{-3}\right.\nonumber \\
 & \left.\left.+\frac{6\pi}{k}\mathcal{U}L_{-1}+\frac{\pi}{k}\left(\mathcal{U}+\frac{12\pi}{k}\tilde{\mathcal{L}}^{2}\right)U_{-1}\right]-i\nu_{IJ}T^{IJ}\right\} d\tau\;,\label{eq:Ebh}
\end{align}
where $\tilde{\mathcal{L}}$ is given by 
\begin{equation}
\tilde{\mathcal{L}}=\mathcal{L}-\frac{5\pi}{2k}{\cal J}_{IJ}{\cal J}^{IJ}\;,
\end{equation}
while the components of the zero modes of the $so\left(M\right)$
Kac-Moody currents, and their corresponding chemical potentials are
constrained to commute by the field equations, and hence 
\begin{equation}
{\cal J}_{\left[I\right.}^{\;K}\nu_{\left.J\right]K}=0\;.
\end{equation}
As above, if we assume that ${\cal J}_{IJ}T^{IJ}$ belong to the Cartan
subalgebra of $so(M)$, this condition implies that the chemical potentials
$\nu_{IJ}T^{IJ}$ also do. Therefore, 
\[
\nu_{2i-1\,2i}=\nu_{i}\epsilon_{2i-1\,2i}\;.
\]

Due to the fact that the thermal cycles are contractible,  the holonomy of the gauge fields along them
has to be trivial. These are the so-called ``regularity conditions". For the branch of solutions that is continuously
connected to the BTZ black hole \cite{BTZ,BHTZ}, possibly endowed
with an $so\left(M\right)$ field, the regularity conditions read
\begin{equation}
e^{a_{\tau}^{sp\left(4\right)}}.e^{i\left(\nu_{IJ}+\frac{5\pi}{k}\xi{\cal J}_{IJ}\right)T^{IJ}}=\left(\begin{array}{cc}
-\mathbbm{1}_{4\times4} & 0\\
0 & \mathbbm{1}_{M\times M}
\end{array}\right)\;.
\end{equation}
Hence, the chemical potentials fulfill
\begin{equation}
\nu_{i}+\frac{5\pi}{k}\xi j_{i}=2\pi n_{i}\;,
\end{equation}
where $n$$_{i}$ stands for a set of integers, and 
\begin{align}
\xi & =\frac{\pi}{5^{2}2}\left[\frac{3^{3}\lambda_{\left[-\right]}^{3}-41\left(3\lambda_{\left[+\right]}-\lambda_{\left[-\right]}\right)\lambda_{\left[-\right]}\lambda_{\left[+\right]}-3^{2}\lambda_{\left[+\right]}^{3}}{\left(\lambda_{\left[-\right]}^{2}-\lambda_{\left[+\right]}^{2}\right)\lambda_{\left[-\right]}\lambda_{\left[+\right]}}\right]\;,\\
\mu & =\frac{3\pi}{5}\left[\frac{3\lambda_{\left[-\right]}-\lambda_{\left[+\right]}}{\left(\lambda_{\left[-\right]}^{2}-\lambda_{\left[+\right]}^{2}\right)\lambda_{\left[-\right]}\lambda_{\left[+\right]}}\right]\;,
\end{align}
with $\lambda_{\left[\pm\right]}$ given by 
\begin{align}
\lambda_{\left[\pm\right]}^{2} & =\frac{10\pi}{k}\left(\mathcal{\tilde{{\cal L}}}\pm\frac{4}{5}\sqrt{\mathcal{\tilde{{\cal L}}}^{2}-\frac{3k}{16\pi}\mathcal{U}}\right)\;.\label{eq:lamdamn}
\end{align} 
One gets exactly the same regularity condition in the $sp(4)$ sector (in terms of $\tilde{\mathcal {L}}$) as in the simple hypergravity case considered in  \cite{Henneaux:2015ywa} (or, for that matter, as in the case of pure $sp(4)$ gravity). 
We also note that the natural value for the integers $n_i$ characterizing the holonomy of the internal $SO(M)$ symmetry is $n_i = 0$ since otherwise there might appear to be a $\delta$-function source of quantized strength in the non-gravitational, internal,  sector, but we shall temporarily allow for more general $n_i$'s to see how these integers enter the entropy.  One could similarly allow for more general solutions of the regularity conditions involving different integers in the $sp(4)$ sector; the above choice corresponds to the BTZ branch.

\subsection{Entropy}

We use the correct canonical expression for the black hole entropy adapted to the above boundary conditions derived first in \cite{PTT1,PTT2}, which 
can also be alternatively written according to \cite{Banados:2012ue,deBoer:2013gz,Bunster:2014mua}
as
\begin{align}
S & =-2k_{4}\text{Im}\left(\text{str}\left[a_{\tau}a_{\varphi}\right]\right)\;\label{EntropyCS}
\end{align}
(in the conventions of \cite{Bunster:2014mua}).
Once evaluated for the solution in (\ref{eq:Ebh}), the black hole
entropy becomes 
\begin{align}
S & =8\pi\mbox{Re}\left[\xi\mathcal{L}+2\mu\mathcal{U}+\frac{1}{2}\nu^{IJ}{\cal J}_{IJ}\right]\ .\label{eq:Seucl}
\end{align}

Plugging then the expressions for the chemical potentials into (\ref{eq:Seucl})
allows one to express the black hole entropy in terms of the (extensive)
global charges. One gets
\begin{equation}
S=\frac{2\pi k}{5}\mbox{Re}\left(3\lambda_{\left[+\right]}+\lambda_{\left[-\right]}+\frac{5}{2k}n_{i}j_{i}\right)\;.\label{eq:BHentropy}
\end{equation}
For the natural $so(M)$ holonomy $n_i=0$, this expression becomes
\begin{equation}
S=\frac{2\pi k}{5}\mbox{Re}\left(3\lambda_{\left[+\right]}+\lambda_{\left[-\right]}\right)\;.\label{eq:BHentropyBis}
\end{equation}
It is only for this branch that the black hole entropy reduces to the horizon area
over $4G$ when the spin-$4$ field is turned off.

Two points are worth being pointed out: (i)  as can already be seen for the coupled pure gauge-gravitational fields without higher spin gauge fields described by the gauge algebra $sl(2, \mathbb{R}) \oplus so(M)$, the $so\left(M\right)$ gauge fields are ``gravitationally
stealth'' in the sense of \cite{cheshireeffect}, i.e., they do not
generate a back reaction on the metric  because their contribution
to the stress energy vanishes; they only contribute to a redefinition of the asymptotic Virasoro generators;
(ii) the black hole entropy of the $n_i =0$ branch (\ref{eq:BHentropyBis})
is also blind to them  if one expresses it in terms of the tilted Virasoro generators but not so if one uses the Virasoro generators fulfilling the above asymptotic algebra and directly related to the mass $M$ and the angular momentum $J$.  The black hole entropy can also detect non-vanishing $n_i$, see (\ref{eq:BHentropy}).

For the $n_i=0$  branch that we consider from now on, the Lorentzian continuation of the entropy reads
\begin{align}
S & =\pi\sqrt{\frac{2}{5}\pi k}\left[\sqrt{\tilde{\mathcal{L}}^{+}}\left(\sqrt{1-\frac{4}{5}\sqrt{1-\frac{3k\mathcal{U}^{+}}{16\pi\left(\tilde{\mathcal{L}}^{+}\right)^{2}}}}+3\sqrt{1+\frac{4}{5}\sqrt{1-\frac{3k\mathcal{U}^{+}}{16\pi\left(\tilde{\mathcal{L}}^{+}\right)^{2}}}}\right)\right.\nonumber \\
 & +\left.\sqrt{\tilde{\mathcal{L}}^{-}}\left(\sqrt{1-\frac{4}{5}\sqrt{1-\frac{3k\mathcal{U}^{-}}{16\pi\left(\tilde{\mathcal{L}}^{-}\right)^{2}}}}+3\sqrt{1+\frac{4}{5}\sqrt{1-\frac{3k\mathcal{U}^{-}}{16\pi\left(\tilde{\mathcal{L}}^{-}\right)^{2}}}}\right)\right]\thinspace.
\end{align}
Requiring the entropy to be
well-defined, i.e., being real and positive, implies that the eigenvalues $\lambda_{\left[\pm\right]}$ should be real. This forces then
the spin-4 charges  to be bounded according to 
\begin{equation}
-\left(\tilde{{\cal L}}^{\pm}\right)^{2}\leq\frac{k}{3\pi}{\cal U}^{\pm}\leq\frac{2^{4}}{3^{2}}\left(\tilde{{\cal L}}^{\pm}\right)^{2}\;,\label{eq:Boundspin4}
\end{equation}
in addition to $\tilde{\mathcal L}^\pm \geq 0$.
The bounds  are saturated in the extremal cases, and only the lower
one in (\ref{eq:Boundspin4}) corresponds to the hypersymmetry bound aforementioned. Note that
the range of positive spin-4 charges is larger than that of the negative
ones.

\section{Killing vector-spinors \label{sec:Killing-vector-spinor-equation}}

Bosonic configurations that admit unbroken hypersymmetries have to
fulfill the following Killing vector-spinor equation 
\begin{equation}
\delta a=d\theta+\left[a,\theta\right]=0\;,
\end{equation}
where the parameter $\theta$ is purely fermionic, given by $\theta=\theta_{I}^{p}{\cal S}_{p}^{I}$
for both copies, and globally well-defined. 

Equivalently, the Killing
vector-spinor equation can be obtained from promoting the corresponding
asymptotic symmetries to hold everywhere and not just asymptotically. Therefore, in the case of
the plus copy ($a_{\varphi}=a_{\varphi}^{+}$), the fermionic parameter
is of the form $\theta=\Omega^{+}\left[0,0,0,\vartheta^{I}\right]$,
which explicitly reads 
\begin{align}
\theta & =-\vartheta_{I}{\cal S}_{\frac{3}{2}}^{I}+\left(\vartheta_{I}^{\prime}+\frac{10}{k}{\cal J}{}_{I}^{\;K}\vartheta_{K}\right){\cal S}_{\frac{1}{2}}^{I}-\frac{1}{2}\left(\vartheta_{I}^{\prime\prime}-\frac{6\pi}{k}\tilde{\mathcal{L}}\vartheta_{I}-\frac{100\pi^{2}}{k^{2}}\mathcal{J}_{IK}\mathcal{J}^{JK}\vartheta_{J}\right.\nonumber \\
 & \left.+\frac{20\pi}{k}\mathcal{J}_{I}^{\;K}\vartheta_{K}^{\prime}\right){\cal S}_{-\frac{1}{2}}^{I}+\frac{1}{6}\left[\left(\vartheta_{I}^{\prime\prime}+\frac{30\pi}{\kappa}\mathcal{J}_{I}^{\;K}\vartheta_{K}^{\prime}-\frac{14\pi}{3k}\tilde{\mathcal{L}}\vartheta_{I}-\frac{300\pi^{2}}{k^{2}}\mathcal{J}_{IK}\mathcal{J}^{JK}\vartheta_{J}\right)^{\prime}\right.\nonumber \\
 & \left.-\frac{140\pi^{2}}{k^{2}}\left(\tilde{\mathcal{L}}\mathcal{J}_{I}^{\;J}+\frac{50\pi}{7k}\mathcal{J}{}_{IM}\mathcal{J}^{MK}\mathcal{J}_{\;K}^{J}\right)\vartheta_{J}\right]{\cal S}_{-\frac{3}{2}}^{I}\;.
\end{align}
The condition that  $a_{\varphi}$ should be left strictly unchanged then implies that the parameters
$\vartheta_{I}$ should satisfy the following differential equations: 
\begin{align}
 & \left[\left(\mathcal{U}+\frac{3\pi}{k}\tilde{{\cal L}}^{2}\right)\delta_{\;I}^{J}+\frac{500\pi^{2}}{3k^{2}}\mathcal{J}_{IK}\left(\tilde{{\cal L}}\mathcal{J}^{JK}+\frac{5\pi}{k}\mathcal{J}_{MN}\mathcal{J}^{JM}\mathcal{J}^{KN}\right)\right]\vartheta_{J}\nonumber \\
 & -\frac{100\pi}{3k}\left(\mathcal{J}_{I}^{\;K}\tilde{{\cal L}}+\frac{10\pi}{k}\mathcal{J}{}_{IM}\mathcal{J}^{ML}\mathcal{J}_{\;L}^{K}\right)\vartheta_{K}^{\prime}\label{eq:KEq1}\\
 & -\frac{5}{3}\left(\delta_{\;I}^{K}\tilde{{\cal L}}+\frac{30\pi}{k}\mathcal{J}_{IJ}\mathcal{J}^{KJ}\right)\vartheta_{K}^{\prime\prime}+\frac{10}{3}\mathcal{J}_{I}^{\;K}\vartheta_{K}^{\prime\prime\prime}+\frac{k}{12\pi}\vartheta_{I}^{\prime\prime\prime\prime}=0\;.\nonumber 
\end{align}

From the experience gathered with black holes within $\left(1,1\right)$
hypergravity or supergravity \cite{Henneaux:2015ywa,Coussaert:1993jp},
it is reasonable to assume that the fermionic parameters are constant,
given by $\vartheta^{I}=\vartheta_{0}^{I}$. The Killing vector-spinor
equations (\ref{eq:KEq1}) then reduce to 
\begin{equation}
\left[\left(\mathcal{U}+\frac{3\pi}{k}\tilde{{\cal L}}^{2}\right)\delta_{\;I}^{J}+\frac{500\pi^{2}}{3k^{2}}\mathcal{J}_{IK}\left(\tilde{{\cal L}}\mathcal{J}^{JK}+\frac{5\pi}{k}\mathcal{J}_{MN}\mathcal{J}^{JM}\mathcal{J}^{KN}\right)\right]\vartheta_{J}=B_{0}^{IK}\vartheta_{K}=0\;,
\end{equation}
which clearly admit non trivial solutions if the matrix $B_{0}^{IK}$ has zero eigenvalues. Since black holes are well-defined
provided $\tilde{\mathcal{L}}\geq0$, and the spin-4 charges fulfill
eq. (\ref{eq:Boundspin4}), the Killing vector-spinor equations (\ref{eq:KEq1})
possess non trivial solutions only when the lower bound in (\ref{eq:Boundspin4})
is saturated, i.e., only for negative spin-4 charges given by 
\[
\mathcal{U}=-\frac{3\pi}{k}(\tilde{\mathcal{L}})^{2}\;.
\]
When this condition is fulfilled, 
\begin{itemize}
\item there is at least one Killing vector-spinor when $M$ is odd (corresponding
to $B_{0}^{2r+1\,2r+1}=0$) and more if some currents $j_{i}$ vanish; 
\item there are Killing vector-spinors when $M$ is even only if some currents
$j_{i}$ vanish. 
\item The maximum number of hypersymmetries is thus $M$.  It is attained when all the $j_i$'s vanish and correspond to $\frac{M}{4}$-hypersymmetry, in agreement with the $\frac14$-hypersymmetry found in  \cite{Henneaux:2015ywa} for $M=1$.
\end{itemize}
It is straightforward to verify that the remaining Killing vector-spinor equation,
that come from preserving the form of the Lagrange multiplier $a_{t}$
globally, is also fulfilled.

\section{Conclusions\label{sec:Conclusions}}

In this note, we have extended the analysis of hypersymmetry bounds of \cite{Henneaux:2015ywa} to extended AdS$_3$ hypergravity.   These bounds follow from the asymptotic symmetry superalgebra and  involve the charges nonlinearly. Just as in \cite{Henneaux:2015ywa}, we have found that the bounds are saturated by a class of extremal black holes, which are hypersymmetric (i.e., possess Killing vector-spinors).  However, not all extremal black holes are hypersymmetric.  The fact that extremality and super/hypersymmetry do not coincide in the context of higher spin black holes has been discussed recently in the thorough work \cite{Banados:2015tft}, which focuses on (an appropriate real form of) the superalgebra $sl(3 \vert 2)$.

Hypersymmetric solutions of a different types (solitons) have been also explored in \cite{Henneaux:2015ywa}.  The extension of that analysis to extended hypergravity is left for future study.

Finally, we note that nonlinear bounds have also been found in the context of asymptotically
flat solutions of hypergravity in the case of fermionic fields of
spin $s=n+\frac{3}{2}$, with $n>0$ (which, in the case of $n=0$,
i.e., supergravity, turn out to be linear) \cite{BoundsHSFlat}.

\section{Acknowledgments}
D.T. holds a Marina Solvay fellowship.
The work of M.H. and D.T. is partially supported by the ERC through
the ``SyDuGraM\textquotedblright \ Advanced Grant, by FNRS-Belgium
(convention FRFC PDR T.1025.14 and convention IISN 4.4503.15) and
by the ``Communauté Française de Belgique\textquotedblright \ through
the ARC program. The work of A.P., D.T. and R.T. is partially funded
by the Fondecyt grants N${^{\circ}}$ 11130262, 11130260, 1130658,
1121031. The Centro de Estudios Cient\'{i}ficos (CECs) is funded by
the Chilean Government through the Centers of Excellence Base Financing
Program of Conicyt.

\begin{appendix}[Explicit form of the asymptotic symmetries]
\label{Appendix:Asy}


The Lie-algebra-valued parameter $\eta^{\pm}$ that appears in the
asymptotic gauge symmetries spanned by $\Omega^{\pm}$ in eq. (\ref{eq:Omegamn})
is given by

\begin{align*}
\eta^{\pm}\left[\epsilon_{\pm},\chi_{\pm},\vartheta_{\pm}^{I}\right] & =-\frac{3\pi}{k}\left(i\psi_{I}^{\pm}\vartheta_{\pm}^{I}+\frac{2}{3}\epsilon_{\pm}\tilde{\mathcal{L}}^{\pm}+2\chi_{\pm}\mathcal{U}^{\pm}-\frac{k}{6\pi}\epsilon_{\pm}^{\prime\prime}\right)L_{\mp1}\mp\epsilon_{\pm}^{\prime}L_{0}\\
 & \hspace{-1.5cm}+\frac{6\pi}{k}\left(\chi_{\pm}\tilde{\mathcal{L}}^{\pm}-\frac{k}{12\pi}\chi_{\pm}^{\prime\prime}\right)U_{\pm1}\mp\frac{2\pi}{k}\left(\chi_{\pm}\tilde{\mathcal{L}}^{\pm\prime}+\frac{8}{3}\chi_{\pm}^{\prime}\tilde{\mathcal{L}}^{\pm}-\frac{k}{12\pi}\chi_{\pm}^{\prime\prime\prime}\right)U_{0}\\
 & \hspace{-1.5cm}-\frac{\pi}{2k}\left[i\psi_{I}^{\pm}\text{\ensuremath{\vartheta}}_{\pm}^{I}+2\left(\mathcal{U}^{\pm}-\frac{1}{2}\tilde{\mathcal{L}}^{\pm\prime\prime}+\frac{12\pi}{k}\left(\tilde{\mathcal{L}}^{\pm}\right)^{2}\right)\chi-\frac{11}{3}\chi_{\pm}^{\prime}\tilde{\mathcal{L}}^{\pm\prime}\right.\\
 & \hspace{-1.5cm}\left.-\frac{14}{3}\chi_{\pm}^{\prime\prime}\tilde{\mathcal{L}}^{\pm}+\frac{k}{12\pi}\chi_{\pm}^{\left(4\right)}\right]U_{\mp1}\pm\chi_{\pm}^{\prime}U_{\pm2}\pm\frac{\pi}{2k}\left[i\psi_{I}^{\pm}\vartheta_{\pm}^{I\prime}+\frac{1}{5}i\psi_{I}^{\pm\prime}\vartheta_{\pm}^{I}\right.\\
 & \hspace{-1.5cm}+\frac{8\pi}{k}i{\cal J}_{\;\;J}^{I\pm}\psi_{I}^{\pm}\vartheta_{\pm}^{J\prime}-\frac{5}{3}\chi_{\pm}^{\prime\prime}\tilde{\mathcal{L}}^{\pm\prime}-\frac{4}{3}\tilde{\mathcal{L}}^{\pm}\chi_{\pm}^{\prime\prime\prime}+\frac{2}{5}\left(\mathcal{U}^{\pm}-\frac{1}{2}\tilde{\mathcal{L}}^{\pm\prime\prime}+\frac{18\pi}{k}\left(\tilde{\mathcal{L}}^{\pm}\right)^{2}\right)^{\prime}\chi_{\pm}\\
 & \hspace{-1.5cm}\left.+\frac{6}{5}\left(\mathcal{U}^{\pm}-\frac{7}{9}\tilde{\mathcal{L}}^{\pm\prime\prime}+\frac{44\pi}{3k}\left(\tilde{\mathcal{L}}^{\pm}\right)^{2}\right)\chi_{\pm}^{\prime}+\frac{k}{60\pi}\chi_{\pm}^{\left(5\right)}\right]U_{\mp2}\\
 & \hspace{-1.5cm}-\frac{\pi}{4k}\left\{ i\psi_{I}^{\pm}\vartheta_{\pm}^{I\prime\prime}+\frac{1}{15}i\left(\psi_{I}^{\pm\prime\prime}-2^{4}\frac{5\pi}{k}\tilde{{\cal \mathcal{L}}}^{\pm}\psi_{I}^{\pm}+2^{3}\frac{5\pi}{k}{\cal J}_{\;\;I}^{J\pm}\psi_{J}^{\pm\prime}+2^{2}\frac{35\pi}{k}{\cal J}_{\;\;I}^{J\pm\prime}\psi_{J}^{\pm}\right.\right.\\
 & \hspace{-1.5cm}\left.-\frac{10^{3}\pi^{2}}{k^{2}}{\cal J}_{\;\;K}^{J\pm}{\cal J}_{I}^{\;\;K\pm}\psi_{J}^{\pm}\right)\vartheta_{\pm}^{I}+\frac{2}{5}i\left(\psi_{I}^{\pm\prime}+\frac{40\pi}{k}{\cal J}_{\;\;I}^{J\pm}\psi_{J}^{\pm\prime}\right)\vartheta_{\pm}^{I\prime}-\chi_{\pm}^{\prime\prime\prime}\tilde{\mathcal{L}}^{\pm\prime}-\frac{4}{5}\epsilon_{\pm}\mathcal{U}^{\pm}\\
 & \hspace{-1.5cm}+\frac{2}{3}\left(\mathcal{U}^{\pm}-\frac{13}{10}\mathcal{L}^{\pm\prime\prime}+\frac{272\pi}{15k}\left(\tilde{\mathcal{L}}^{\pm}\right)^{2}\right)\chi_{\pm}^{\prime\prime}+\frac{8}{15}\left(\mathcal{U}^{\pm}-\frac{17}{24}\tilde{\mathcal{L}}^{\pm\prime\prime}+\frac{241\pi}{12k}\left(\tilde{\mathcal{L}}^{\pm}\right)^{2}\right)^{\prime}\chi_{\pm}^{\prime}\\
 & \hspace{-1.5cm}+\frac{40\pi}{3k}\left[i\psi_{I}^{\pm}\psi^{\pm I\prime}-\frac{12\pi}{5k}\left(\tilde{\mathcal{L}}^{\pm}\right)^{3}-\frac{11}{5^{2}}\mathcal{U}^{\pm}\tilde{\mathcal{L}}^{\pm}+\frac{3^{2}}{5^{2}}\left(\tilde{\mathcal{L}}^{\pm\prime}\right)^{2}+\frac{23}{50}\tilde{\mathcal{L}}^{\pm\prime\prime}\tilde{\mathcal{L}}^{\pm}\right.\\
 & \hspace{-1.5cm}\left.+\frac{k}{10^{2}\pi}\left(\mathcal{U}^{\pm}-\frac{1}{2}\mathcal{\tilde{L}}^{\pm\prime\prime}\right)^{\prime\prime}+\frac{10\pi}{k}i{\cal J}^{IJ\pm}\psi_{I}^{\pm}\psi_{J}^{\pm}\right]\chi_{\pm}\left.-\frac{5}{9}\chi_{\pm}^{\left(4\right)}\mathcal{L}^{\pm}+\frac{k}{180\pi}\chi_{\pm}^{\left(6\right)}\right\} U_{\mp3}\\
 & \hspace{-1.5cm}-\frac{2\pi}{k}\left\{ \epsilon_{\pm}\psi_{I}^{\pm}+\frac{1}{2}\vartheta_{\pm I}\mathcal{L}^{\pm\prime}+\frac{7}{6}\left[\delta_{\;I}^{J}\mathcal{L}^{\pm}-\frac{15}{7}\left({\cal J}_{I}^{\;\;J\pm\prime}-\frac{10}{k}{\cal J}^{JK\pm}{\cal J}_{IK}^{\pm}\right)\right]\vartheta_{\pm J}^{\prime}\right.\\
 & \hspace{-1.5cm}-\frac{5}{2}{\cal J}_{I}^{\;\;J\pm}\vartheta_{\pm J}^{\prime\prime}-\frac{5}{3}\left[\psi_{I}^{\pm\prime\prime}-\frac{52\pi}{5k}\left(\tilde{\mathcal{L}}^{\pm}\psi_{I}^{\pm}-\frac{25}{13}{\cal J}_{I}^{\;\;J\pm}\psi_{J}^{\pm\prime}-\frac{25}{26}{\cal J}_{I}^{\;\;J\pm\prime}\psi_{J}^{\pm}\right.\right.\\
 & \hspace{-1.5cm}\left.\left.+\frac{125}{13}\psi_{P}^{\pm}{\cal J}^{KP\pm}{\cal J}_{KI}^{\pm}\right)\right]\chi_{\pm}-\frac{25}{6}\left(\psi_{I}^{\pm\prime}+\frac{10\pi}{k}\psi_{J}^{\pm}{\cal J}_{I}^{\;\;J\pm}\right)\chi_{\pm}^{\prime}\\
 & \hspace{-1.5cm}-\frac{5}{6}\left[{\cal J}_{IJ}^{\pm\prime\prime}-\frac{10\pi}{k}\left({\cal J}_{I}^{\;\;K\pm\prime}{\cal J}_{JK}^{\pm}+\frac{7}{5}{\cal \tilde{L}}{\cal J}_{IJ}^{\pm}+2{\cal J}_{I}^{\;\;K\pm}{\cal J}_{JK}^{\pm\prime}+\frac{10\pi}{k}{\cal J}^{MN\pm}{\cal J}_{IM}^{\pm}{\cal J}_{JN}^{\pm}\right)\right]\vartheta_{\pm}^{J}\\
 & \hspace{-1.5cm}\left.-\frac{17}{6}\chi_{\pm}^{\prime\prime}\psi_{I}^{\pm}-\frac{k}{12\pi}\vartheta_{\pm I}^{\prime\prime\prime}\right\} {\cal S}_{\mp\frac{3}{2}}^{I}\pm\frac{3\pi}{k}\left[\left(\delta_{\;I}^{J}\tilde{\mathcal{L}}^{\pm}-\frac{5}{3}{\cal J}_{I}^{\;\;J\pm\prime}+\frac{50}{3k}{\cal J}^{JK\pm}{\cal J}_{IK}^{\pm}\right)\vartheta_{\pm J}\right.\\
 & \hspace{-1.5cm}\left.-5\chi_{\pm}^{\prime}\psi_{I}^{\pm}-\frac{10}{3}{\cal J}_{I}^{\;\;J\pm}\vartheta_{\pm J}^{\prime}-\frac{10}{3}\left(\psi_{I}^{\pm\prime}+\frac{10\pi}{k}\psi_{J}^{\pm}{\cal J}_{I}^{\;\;J\pm}\right)\chi_{\pm}-\frac{k}{6\pi}\vartheta_{\pm I}^{\prime\prime}\right]{\cal S}_{\mp\frac{1}{2}}^{I}\\
 & \hspace{-1.5cm}+\frac{20\pi}{k}\left(\chi_{\pm}\psi_{I}^{\pm}+\frac{1}{2}{\cal J}_{IJ}^{\pm}\vartheta_{\pm}^{J}+\frac{k}{20\pi}\vartheta_{\pm I}^{\prime}\right){\cal S}_{\pm\frac{1}{2}}^{I}\thinspace,
\end{align*}
while the transformation laws of the fields ${\cal L}^{\pm}$, ${\cal U}^{\pm}$,
$\mathcal{J}_{IJ}^{\pm}$, $\psi^{\pm}$, explicitly reads
\begin{align*}
\mathcal{\delta L}^{\pm} & =2\epsilon_{\pm}^{\prime}\mathcal{L}^{\pm}+\epsilon_{\pm}\mathcal{L}^{\pm\prime}-\frac{k}{4\pi}\epsilon_{\pm}^{\prime\prime\prime}+3\mathcal{U}^{\pm\prime}\chi_{\pm}+4\mathcal{U}^{\pm}\chi_{\pm}^{\prime}+\frac{5}{2}i\psi_{I}^{\pm}\left(\vartheta_{\pm}^{I}\right)^{\prime}+\frac{3}{2}i\psi_{I}^{\pm\prime}\vartheta_{\pm}^{I}-{\cal J}_{IJ}^{\pm}\left(\zeta_{\pm}^{IJ}\right)^{\prime}\;,\\
\delta j_{IJ}^{\pm} & =\epsilon_{\pm}^{\prime}j_{IJ}^{\pm}+\epsilon_{\pm}j_{IJ}^{\pm\prime}+4{\cal J}_{\left[I\right|}^{\pm K}\zeta_{\left.J\right]K}^{\pm}+\frac{k}{5\pi}\zeta_{IJ}^{\pm\prime}-2i\psi_{\left[I\right|}^{\pm}\vartheta_{\pm\left|J\right]}\;,
\end{align*}

\begin{align*}
\delta\psi_{I}^{\pm} & =\frac{5}{2}\epsilon_{\pm}^{\prime}\psi_{I}^{\pm}+\epsilon_{\pm}\psi_{I}^{\pm\prime}+2\zeta_{I}^{\;K}\psi_{K}^{\pm}-\left[\left(\mathcal{U}^{\pm}-\frac{1}{2}\tilde{\mathcal{L}}^{\pm\prime\prime}\right)\delta_{\;I}^{K}+\frac{5}{6}\left({\cal J}_{I}^{\pm K}\right)^{\prime\prime\prime}+\frac{3\pi}{k}\Lambda_{\pm I}^{\left(4\right)K}\right]\vartheta_{\pm K}\\
 & +\frac{5}{3}\left[\left(\tilde{\mathcal{L}}^{\pm}\delta_{\;I}^{K}-2\left({\cal J}_{I}^{\pm K}\right)^{\prime}\right)\vartheta_{\pm K}^{\prime}-\frac{k}{20\pi}\vartheta_{\pm I}^{\prime\prime\prime}\right]^{\prime}+\frac{82\pi}{3k}\left(\Lambda_{\pm}^{\left(9/2\right)\prime}-\frac{23}{82}\Lambda_{\text{\ensuremath{\pm}}}^{\left(11/2\right)}-\frac{5k}{82\pi}\psi_{I}^{\pm\prime\prime\prime}\right)\chi_{\pm}\\
 & -7\left(\psi_{I}^{\pm\prime}+\frac{10\pi}{k}\Lambda_{\text{\ensuremath{\pm}}I}^{\left(7/2\right)}\right)\chi_{\pm}^{\prime\prime}+\frac{35\pi}{k}\left(\Lambda_{\pm I}^{\left(9/2\right)}-\frac{k}{6\pi}\psi_{I}^{\pm\prime\prime}\right)\chi_{\pm}^{\prime}-\frac{35}{12}\chi_{\pm}^{\prime\prime\prime}\psi_{I}^{\pm}\\
 & -\frac{5}{3}\left[\left({\cal J}_{I}^{\pm K}\right)^{\prime}+\frac{30\pi}{k}\Lambda_{\text{\ensuremath{\pm}}I}^{\left(2\right)K}\right]\vartheta_{\pm K}^{\prime\prime}-\frac{10}{3}{\cal J}_{\pm I}^{\;\;K}\vartheta_{\pm K}^{\prime\prime\prime}+\frac{10^{2}}{3k}\Lambda_{\text{\ensuremath{\pm}}I}^{\left(3\right)K}\vartheta_{\pm K}^{\prime}\;,
\end{align*}
\begin{align*}
\delta\mathcal{U}^{\pm} & =4\epsilon_{\pm}^{\prime}\mathcal{U}^{\pm}+\epsilon_{\pm}\mathcal{U}^{\pm\prime}+\frac{23\pi}{3k}i\left(\Lambda_{\pm I}^{\left(11/2\right)}+\Lambda_{\pm I}^{\left(9/2\right)\prime}+\frac{210}{23}\Lambda_{\pm I}^{\left(7/2\right)\prime\prime}-\frac{k}{92\pi}\psi_{I}^{\pm\prime\prime\prime}\right)\vartheta_{\pm}^{I}\\
 & +\frac{35\pi}{k}i\left(\Lambda_{\pm I}^{\left(9/2\right)}+4\Lambda_{\pm I}^{\left(7/2\right)\prime}-\frac{k}{60\pi}\psi_{I}^{\pm\prime\prime}\right)\left(\vartheta_{\pm}^{I}\right)^{\prime}-\frac{7}{4}i\left(\psi_{I}^{\pm\prime}-\frac{40\pi}{k}\Lambda_{\pm I}^{\left(7/2\right)}\right)\left(\vartheta_{\pm}^{I}\right)^{\prime\prime}\\
 & -\frac{35}{12}i\psi_{I}^{\pm}\left(\vartheta_{\pm}^{I}\right)^{\prime\prime\prime}-\frac{1}{6}\left[\left(\mathcal{U}^{\pm}-\frac{1}{2}\mathcal{\tilde{L}}^{\pm\prime\prime}\right)^{\prime\prime}+\frac{144}{k}\left(\Lambda_{\pm}^{\left(6\right)}-\frac{49}{216}\tilde{\Lambda}_{\pm}^{\left(4\right)\prime\prime}\right)\right]^{\prime}\chi_{\pm}\\
 & -\frac{5}{6}\left[\left(\mathcal{U}^{\pm}-\frac{2}{3}\mathcal{\tilde{L}}^{\pm\prime\prime}\right)^{\prime\prime}+\frac{288}{5k}\Lambda_{\pm}^{\left(6\right)}\right]\chi_{\pm}^{\prime}+\frac{14}{9}\left[\mathcal{\tilde{L}}^{\pm\prime\prime}-\frac{27}{28}\mathcal{U}^{\pm}-\frac{21\pi}{k}\tilde{\Lambda}_{\pm}^{\left(4\right)}\right]^{\prime}\chi_{\pm}^{\prime\prime}\\
 & +\frac{7}{3}\left[\mathcal{\tilde{L}}^{\pm\prime\prime}-\frac{3}{7}\mathcal{U}^{\pm}-\frac{28\pi}{3k}\tilde{\Lambda}_{\pm}^{\left(4\right)}\right]\chi^{\prime\prime\prime}+\frac{35}{18}\tilde{\mathcal{L}}^{\pm\prime}\chi_{\pm}^{\left(4\right)}+\frac{7}{9}\tilde{\mathcal{L}}^{\pm}\chi_{\pm}^{\left(5\right)}-\frac{k}{2^{4}3^{2}\pi}\chi_{\pm}^{\left(7\right)}\;,
\end{align*}
where
\begin{align*}
\Lambda_{\pm IJ}^{\left(2\right)} & ={\cal J}_{IK}^{\pm}{\cal J}_{\;\;\;J}^{\pm K}\;,\\
\Lambda_{\pm IJ}^{\left(3\right)} & =\tilde{{\cal L}}^{\pm}\mathcal{J}_{IJ}^{\pm}-\Lambda_{\pm IJ}^{\left(2\right)\prime}-\mathcal{J}_{IK}^{\pm}\left(\mathcal{J}_{\;\;\;J}^{\pm K}\right)^{\prime}-\frac{10\pi}{k}\Lambda_{\pm I}^{\left(2\right)K}\mathcal{J}_{KJ}^{\pm}\;,\\
\Lambda_{\pm IJ}^{\left(4\right)} & =\delta_{IJ}\left(\tilde{{\cal L}}^{\pm}\right)^{2}-\frac{50}{9}\left[{\cal J}_{IJ}^{\pm}{\cal \tilde{L}}^{\pm}-\frac{3}{2}\Lambda_{\pm IJ}^{\left(2\right)\prime}+{\cal J}_{IK}^{\pm\prime}{\cal J}_{\;\;\;J}^{\pm K}\right]^{\prime}-\frac{25}{9}{\cal J}_{IK}^{\pm\prime}{\cal J}_{\;\;\;J}^{\pm K\prime}\\
 & -\frac{500\pi}{9k}\left[{\cal \tilde{L}}^{\pm}\Lambda_{\pm IJ}^{\left(2\right)}-\frac{1}{2}{\cal J}^{\pm MN}\left({\cal J}_{IM}^{\pm}{\cal J}_{NJ}^{\pm}\right)^{\prime}-{\cal J}_{IM}^{\pm}\left({\cal J}^{\pm MN}{\cal J}_{NJ}^{\pm}\right)^{\prime}\right]\\
 & +\frac{2500\pi}{9k^{2}}\Lambda_{\pm I}^{\left(2\right)K}\Lambda_{\pm KJ}^{\left(2\right)}\;,\\
\tilde{\Lambda}_{\pm}^{\left(4\right)} & =\left(\tilde{{\cal L}}^{\pm}\right)^{2}\;,
\end{align*}
 
\begin{align*}
\Lambda_{\pm}^{\left(6\right)} & =-\frac{7}{18}\mathcal{U}^{\pm}\mathcal{\tilde{L}}^{\pm}+\frac{295}{432}\left(\mathcal{\tilde{L}}^{\pm\prime}\right)^{2}+\frac{22}{27}\mathcal{\tilde{L}}^{\pm\prime\prime}\mathcal{\tilde{L}}^{\pm}+\frac{25}{12}i\psi_{K}^{\pm}\psi^{\pm K\prime}\\
 & -\frac{8\pi}{3k}\left[\left(\mathcal{\tilde{L}}^{\pm}\right)^{3}-\frac{125}{16}i{\cal J}^{\pm KL}\psi_{K}^{\pm}\psi_{L}^{\pm}\right]\;.\\
\Lambda_{\pm I}^{\left(7/2\right)} & =\psi_{K}^{\pm}{\cal J}_{I}^{\pm K}\;,\\
\Lambda_{\pm I}^{\left(9/2\right)} & =\mathcal{\tilde{L}}^{\pm}\psi_{I}^{\pm}+\frac{5}{3}\psi_{K}^{\pm}{\cal J}_{I}^{\pm K\prime}-\frac{10}{3}\Lambda_{\pm I}^{\left(7/2\right)\prime}-\frac{50\pi}{3k}\psi_{K}^{\pm}\Lambda_{\pm I}^{\left(2\right)K}\;,\\
\Lambda_{\pm I}^{\left(11/2\right)} & =\frac{27}{23}\left(\tilde{\mathcal{L}}^{\pm\prime}\psi_{I}^{\pm}-\frac{260}{27}{\cal J}_{I}^{\pm K\prime}\psi_{K}^{\pm\prime}-\frac{370}{81}{\cal J}_{I}^{\pm K\prime}\psi_{K}^{\pm\prime\prime}-\frac{260}{81}{\cal J}_{I}^{\pm K\prime\prime}\psi_{K}^{\pm}\right)\\
 & +\frac{820\pi}{23k}\left(\tilde{\mathcal{L}}^{\pm}\Lambda_{\pm I}^{\left(7/2\right)}-\frac{25}{41}\Lambda_{\pm I}^{\left(2\right)K}\psi_{K}^{\pm\prime}+\frac{130}{123}\Lambda_{\pm K}^{\left(7/2\right)}{\cal J}_{I}^{\pm K\prime}+\frac{55}{123}{\cal J}_{I}^{\pm K}\Lambda_{\pm K}^{\left(7/2\right)\prime}\right)\\
 & +\frac{5000\pi^{2}}{23k^{2}}\Lambda_{\pm K}^{\left(7/2\right)}\Lambda_{\pm I}^{\left(2\right)K}\;.
\end{align*}
Here the prime denotes derivative with respect to $\varphi$, and $\chi^{(n)}_\pm$ denotes the $n$-th derivative of $\chi_\pm$.

\end{appendix}

\begin{appendix}[Poisson brackets of the canonical generators]

\label{Appendix:PB}

The Poisson brackets of the asymptotic symmetry generators are given
by 
\begin{align}
\left[{\cal L}\left(\varphi\right),{\cal L}\left(\phi\right)\right]_{PB} & =-2\delta^{\prime}\left(\varphi-\phi\right)\mathcal{L}\left(\varphi\right)-\delta\left(\varphi-\phi\right)\mathcal{L}^{\prime}\left(\varphi\right)+\frac{k}{4\pi}\delta^{\prime\prime\prime}\left(\varphi-\phi\right)\;,\nonumber \\
\left[{\cal L}\left(\varphi\right),{\cal U}\left(\phi\right)\right]_{PB} & =-4\delta^{\prime}\left(\varphi-\phi\right)\mathcal{U}\left(\varphi\right)-3\delta\left(\varphi-\phi\right)\mathcal{U}^{\prime}\left(\varphi\right)\;,\nonumber \\
\left[{\cal L}\left(\varphi\right),{\cal J}^{IJ}\left(\phi\right)\right]_{PB} & =-{\cal J}^{IJ}\left(\varphi\right)\delta^{\prime}\left(\varphi-\phi\right)\;,\nonumber \\
\left[{\cal L}\left(\varphi\right),\psi^{\left[I\right]}\left(\phi\right)\right]_{PB} & =-\frac{5}{2}\delta^{\prime}\left(\varphi-\phi\right)\psi^{\left[I\right]}\left(\varphi\right)-\frac{3}{2}\delta\left(\varphi-\phi\right)\psi^{\left[I\right]\prime}\left(\varphi\right)\;,\nonumber \\
\left[{\cal U}\left(\varphi\right),{\cal J}^{IJ}\left(\phi\right)\right]_{PB} & =0\;,\nonumber \\
\left[{\cal J}^{IJ}\left(\varphi\right),{\cal J}_{KL}\left(\phi\right)\right]_{PB} & =-4\delta_{\;\left[K\right.}^{\left[I\right.}{\cal J}_{\;\left.L\right]}^{\left.J\right]}\left(\varphi\right)\delta\left(\varphi-\phi\right)-\frac{k}{5\pi}\delta_{\;\left[K\right.}^{\left[I\right.}\delta_{\;\left.L\right]}^{\left.J\right]}\delta^{\prime}\left(\varphi-\phi\right)\;,\\
\left[{\cal J}^{IJ}\left(\varphi\right),\psi^{K}\left(\phi\right)\right]_{PB} & =2\delta^{K\left[I\right.}\psi^{\left.J\right]}\left(\varphi\right)\delta\left(\varphi-\phi\right)\;,\nonumber \\
\left[{\cal U}\left(\varphi\right),\psi^{I}\left(\phi\right)\right]_{PB} & =\frac{1}{12}\left\{ \left[\psi^{I\prime\prime}\left(\varphi\right)-\frac{92\pi}{k}\left(\Lambda^{\left(9/2\right)I}\left(\varphi\right)+\frac{210}{23}\Lambda^{\left(7/2\right)I\prime}\left(\varphi\right)\right)\right]^{\prime}\right.\nonumber \\
 & \left.-\frac{92\pi}{k}\Lambda^{\left(11/2\right)I}\left(\varphi\right)\right\} \delta\left(\varphi-\phi\right)+\frac{7}{12}\left(\psi^{I\prime\prime}\left(\varphi\right)-\frac{60\pi}{k}\Lambda^{\left(9/2\right)I}\left(\varphi\right)\right.\nonumber \\
 & \left.-\frac{240\pi}{k}\Lambda^{\left(7/2\right)I\prime}\left(\varphi\right)\right)\delta^{\prime}\left(\varphi-\phi\right)+\frac{35}{12}\psi^{I}\left(\varphi\right)\delta^{\prime\prime\prime}\left(\varphi-\phi\right)\nonumber \\
 & +\frac{7}{4}\left(\psi^{I\prime}-\frac{40\pi}{k}\Lambda^{\left(7/2\right)I}\left(\varphi\right)\right)\delta^{\prime\prime}\left(\varphi-\phi\right)\;,\nonumber 
\end{align}
\begin{align}
\left[{\cal U}\left(\varphi\right),{\cal U}\left(\phi\right)\right]_{PB} & =\frac{5}{6}\left[\left(\mathcal{U}\left(\varphi\right)-\frac{2}{3}\mathcal{\tilde{L}}^{\prime\prime}\left(\varphi\right)\right)^{\prime\prime}+\frac{288\pi}{5k}\Lambda^{\left(6\right)}\left(\varphi\right)\right]\delta^{\prime}\left(\varphi-\phi\right)\nonumber \\
 & +\frac{1}{6}\left[\left(\mathcal{U}\left(\varphi\right)-\frac{1}{2}\mathcal{\tilde{L}}^{\prime\prime}\left(\varphi\right)-\frac{98\pi}{3k}\tilde{\Lambda}^{\left(4\right)}\left(\varphi\right)\right)^{\prime\prime}+\frac{144\pi}{k}\Lambda^{\left(6\right)}\left(\varphi\right)\right]^{\prime}\delta\left(\varphi-\phi\right)\nonumber \\
 & +\frac{3}{2}\left(\mathcal{U}\left(\varphi\right)-\frac{28}{27}\tilde{\mathcal{L}}^{\prime\prime}\left(\varphi\right)+\frac{196\pi}{9k}\tilde{\Lambda}^{\left(4\right)}\left(\varphi\right)\right)^{\prime}\delta^{\prime\prime}\left(\varphi-\phi\right)\nonumber \\
 & +\left(\mathcal{U}\left(\varphi\right)-\frac{7}{3}\tilde{\mathcal{L}}^{\prime\prime}\left(\varphi\right)+\frac{196\pi}{9k}\tilde{\Lambda}^{\left(4\right)}\left(\varphi\right)\right)\delta^{\prime\prime\prime}\left(\varphi-\phi\right)\nonumber \\
 & -\frac{35}{18}\tilde{\mathcal{L}}^{\prime}\left(\varphi\right)\delta^{\left(4\right)}\left(\varphi-\phi\right)-\frac{7}{9}\mathcal{\tilde{L}}\left(\varphi\right)\delta^{\left(5\right)}\left(\varphi-\phi\right)+\frac{k}{144\pi}\delta^{\left(7\right)}\left(\varphi-\phi\right)\;,\nonumber \\
i\left[\psi^{I}\left(\varphi\right),\psi^{J}\left(\phi\right)\right]_{PB} & =\delta\left(\phi-\varphi\right)\left[\delta^{IJ}\mathcal{U}\left(\varphi\right)-\frac{1}{2}\delta^{IJ}\mathcal{\tilde{L}}^{\prime\prime}\left(\varphi\right)+\frac{5}{6}\mathcal{J}^{IJ\prime\prime}\left(\varphi\right)+\frac{3\pi}{k}\Lambda^{\left(4\right)\left(IJ\right)}\left(\varphi\right)\right.\nonumber \\
 & \left.+\frac{25\pi}{k}\Lambda^{\left(2\right)JI\prime}\left(\varphi\right)+\frac{50\pi}{3k}\Lambda^{\left(3\right)JI\prime}\left(\varphi\right)\right]+\frac{10}{3}\mathcal{J}^{IJ}\left(\varphi\right)\text{\ensuremath{\delta}}^{\prime\prime\prime}\left(\phi-\varphi\right)\nonumber \\
 & +\delta^{\prime}\left(\phi-\varphi\right)\left[\frac{10}{3}\mathcal{J}^{IJ\prime\prime}\left(\varphi\right)-\frac{5}{3}\delta^{IJ}\mathcal{\tilde{L}}^{\prime}\left(\varphi\right)-\frac{100\pi}{3k}\Lambda^{\left(3\right)\left[IJ\right]}\left(\varphi\right)\right.\nonumber \\
 & \left.+\frac{50\pi}{k}\Lambda^{\left(2\right)JI\prime}\left(\varphi\right)\right]+\frac{k}{12\pi}\delta^{IJ}\delta^{\prime\prime\prime\prime}\left(\phi-\varphi\right)\nonumber \\
 & +\delta^{\prime\prime}\left(\phi-\varphi\right)\left[5\mathcal{J}^{IJ\prime}\left(\varphi\right)-\frac{5}{3}\delta^{IJ}\tilde{\mathcal{L}}\left(\varphi\right)+\frac{50\pi}{k}\Lambda^{\left(2\right)IJ}\left(\varphi\right)\right]\;,
\end{align}
so that once expanded in Fourier modes, the algebra corresponds to
the one in eqs. (\ref{eq:PBFermionic})

\end{appendix}

\end{document}